\documentclass[prc,showpacs,floatfix]{revtex4}
\usepackage{bm}
\usepackage{graphicx}
\newcommand{\beq}{\begin{equation}}
\newcommand{\eeq}{\end{equation}}
\newcommand{\beqa}{\begin{eqnarray}}
\newcommand{\eeqa}{\end{eqnarray}}
\newcommand{\bra}[1]{\mbox{$\langle #1|$}}
\newcommand{\ket}[1]{\mbox{$|#1\rangle$}}
\begin{document}
\title{
\hfill{\small {\bf MKPH-T-02-06}}\\
{ \bf Influence of final state interaction on incoherent pion
photoproduction on the deuteron in the region of the
$\Delta$-resonance}}

\author{E.M.\ Darwish\footnote{Present address: Physics Department,
Faculty of Science, South Valley University, Sohag, Egypt.}, 
H.\ Arenh\"ovel, and M.\ Schwamb}
\affiliation{
Institut f\"{u}r Kernphysik, Johannes Gutenberg-Universit\"{a}t,
       D-55099 Mainz, Germany}
\date{\today}
\begin{abstract} 
The influence of final state $NN$- and $\pi N$-rescattering in
incoherent pion photoproduction on the deuteron has been 
investigated. For the elementary photoproduction operator an effective
Lagrangian model is used which describes well the elementary
reaction. The interactions in the final two-body subsystems
are taken in separable form. While $NN$-rescattering shows quite a
significant effect, particularly strong for neutral pion production,
$\pi N$-rescattering is almost negligible. Inclusion of such effects
leads to an improved and quite satisfactory agreement with experiment. 
\end{abstract}

\maketitle

\section{Introduction}
\label{sec1}
The particular interest in pion photoproduction on the deuteron lies
in the fact that the simple and well known deuteron structure allows
one to obtain information on the production process on the neutron
which otherwise is difficult to obtain in view of the absence of any
free neutron targets. The essential idea behind this reasoning is that
for quasifree kinematics the dominant production process is given by
the elementary reaction on one nucleon while the other acts merely as
a spectator. 
However, this is possible only if competing two-body processes like
final state interaction (FSI) in the $\pi NN$ system and possible two-body
exchange current contributions are under control. 

Early studies of this reaction are done 
in~\cite{ChL51,LaF52,BlL77}. Approximate treatments of final state
interaction effects within a diagrammatic approach have been reported 
in~\cite{Lag78,Lag81}. In that work, a
comparison with experimental data was possible only for $\pi^-$
production~\cite{Be+73}, and a satisfactory agreement was found. The  
authors noted that the FSI effects are quite small for the charged
pion photoproduction reactions in comparison to the neutral
channel. More recently Levchuk {\it et al.} \cite{LeP96} studied 
quasifree $\pi^0$ photoproduction on the neutron via the 
$d(\gamma,\pi^0)np$ reaction using the
elementary photoproduction operator of Blomqvist and Laget 
\cite{BlL77}. In agreement with the results of~\cite{Lag81}, they
found that the largest rescattering effects arise from the $np$
final state interaction leading to a strong
reduction of the cross section at pion forward angles, but are much less
important in backward direction. The experimental data from
\cite{Kru99} for the $d(\gamma,\pi^0)np$ reaction qualitatively
support this prediction although a direct comparison was not possible. 
The threshold region was explored in~\cite{Le+00} where a sizeable 
effect from $\pi N$ rescattering was noted via intermediate charged
pion production with subsequent charge exchange. 
Recently, Levchuk {\it et al.}~\cite{LeS00}
modified the theoretical predictions of~\cite{LeP96} using a more
realistic elementary production operator and including also the
charged pion production channels but considering only $NN$ rescattering 
for which the Bonn r-space potential~\cite{MaH87} was used. 
The elementary production operator
was taken from the SAID~\cite{Said} and MAID~\cite{Maid} multipole
analyses. The sizeable effects from $NN$ FSI were confirmed and good agreement
with the experimental data was achieved. 

The present paper is a natural extension of our work in~\cite{ScA96} where
this process was studied in the pure impulse approximation (IA), i.e., 
without inclusion of any FSI or two-body currents. First of all, we
were interested in the question whether inclusion of FSI would lead to
a good description of the available data, in particular with respect to
the recent data on incoherent $\pi^0$ production on the
deuteron~\cite{Kru99}. Although quite a good description was already
achieved in~\cite{LeS00}, we were puzzeled by the fact 
that the results for the IA of this work showed certain significant
differences to our IA results~\cite{ScA96} for charged pion production,
which is most obvious in the differential cross sections at forward
angles. The origin of this discrepancy was not clear. Furthermore, 
it was an open question whether the inclusion of rescattering contributions
would lead to a different result. Therefore, we have included 
in the present work as a first step the
presumably must important part of the FSI, namely the full
hadronic rescattering in all two-body subsystems of the final
state, i.e., $NN$- and $\pi N$-rescattering, whereas the third
particle is treated as a spectator. It is still an approximate
treatment, the same as in~\cite{LeS00}, insofar as 
only the complete scattering in either the $NN$- or the $\pi N$
subsystems are considered, and not a genuine three-body approach. In
particular, it will remain a future task to see how critical the
violation of unitarity will be. 

In the next section we will briefly review the model for the
elementary photoproduction amplitude which will serve as an input for
the reaction on the deuteron. Sect.~\ref{sec3} will introduce the
general form of the differential cross section for incoherent pion
photoproduction on the deuteron. The separate contributions of the
impulse approximation and the two rescatterings to the transition
matrix are described in Sect.~\ref{sec4}. Details of the actual
calculation and the results are presented and discussed in
Sect.~\ref{sec5}. Finally, we close in Sect.~\ref{sec6} with a summary
and an outlook.  

\section{The elementary pion photoproduction on the nucleon}
\label{sec2}
\begin{figure}[htb]
\includegraphics[scale=.7]{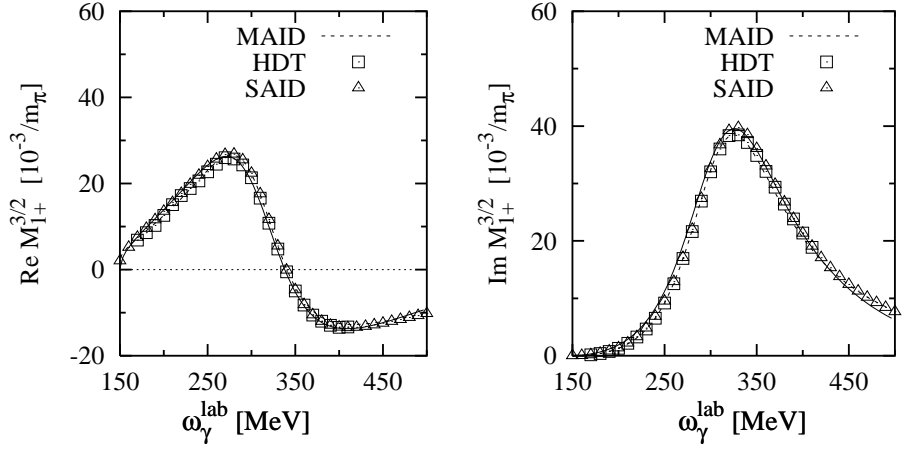}
\caption{Real and imaginary parts of the $M_{1+}^{3/2}$
multipole. Notation: solid curves: present model; short-dashed curves: 
MAID~\protect\cite{Maid}. Data points: from \protect\cite{Said}
(SAID, solution: September 2000), \protect\cite{HaD98} (HDT).}
\label{multipoles}
\end{figure}

For the elementary photoproduction operator, we have taken 
the effective Lagrangian model of Schmidt {\it et al.}~\cite{ScA96}
since it is given in an arbitrary frame of reference and 
allows a well defined off-shell continuation as required for
studying pion production on nuclei.
It is in contrast to other approaches, where the elementary amplitude 
is constructed first on-shell in the photon-nucleon c.m.\ frame with 
subsequent boost into an arbitrary reference frame and some prescription 
for the off-shell continuation. In the latter method, one loses terms which 
by chance vanish in the c.m.\ frame \cite{BrA97}. In our approach,
the only uncertainty arises from the assignment of the invariant energy
for the photon-nucleon subsystem in the resonance propagators as has 
been discussed in detail in \cite{BrA97}. Here we use the spectator 
on-shell approach. The model of~\cite{ScA96} consists of the standard 
pseudovector Born terms and the contribution of the $\Delta(1232)$ 
resonance. For details we refer to~\cite{ScA96}. 
The parameters of the $\Delta$ resonance are fixed by
fitting the experimental $M_{1+}^{3/2}$ multipole. With respect to the
parameters used in~\cite{ScA96}, there was only a
slight change in the mass of the $\Delta(1232)$ resonance for which we
took a value of 1233 MeV. The quality of the model can be judged by a
comparison with the MAID analysis \cite{Maid}, the Mainz dispersion
analysis~\cite{HaD98} and the VPI analysis \cite{Said} as shown in 
Fig.~\ref{multipoles}, and one notes quite a good agreement. 

\begin{figure}[htb]
\includegraphics[scale=.7]{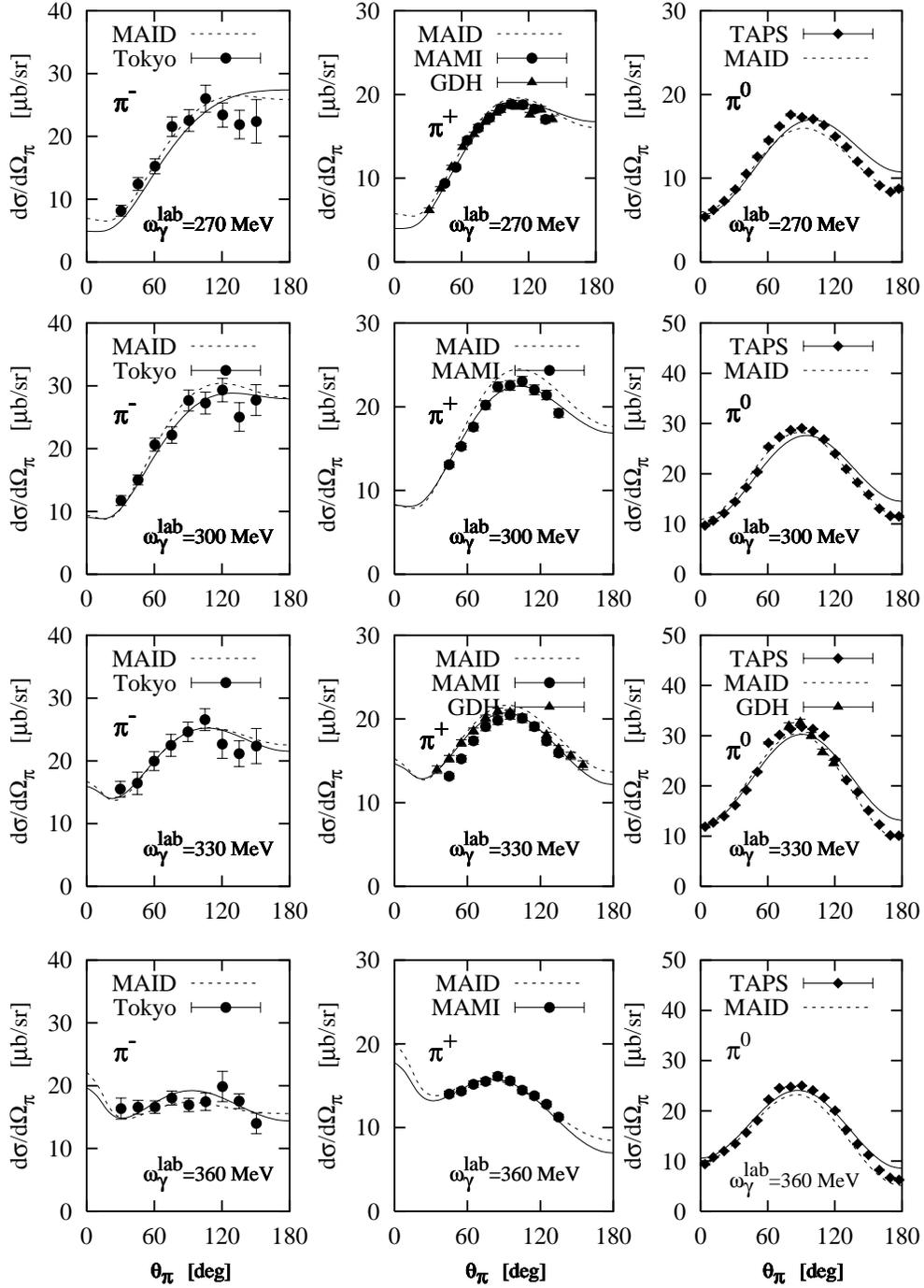}
\caption{Differential cross section for the elementary reaction on 
the nucleon for the three charge states of the pion at various  
photon energies. Left panels: $\pi^-$, middel panels: $\pi^+$, and 
right panels: $\pi^0$. The solid curves: present model; short-dashed curves: 
MAID~\protect\cite{Maid}. Experimental data from
\protect\cite{Fu+77} (Tokyo) for $\pi^-$, \protect\cite{Leu00} (MAMI),
\protect\cite{Pre00} (GDH) for $\pi^0$, and \protect\cite{Be+97} (TAPS),
\protect\cite{Pre00} (GDH) for $\pi^+$.}  
\label{gnucdiff}
\end{figure}

In Fig.~\ref{gnucdiff} we compare our results 
for the differential cross sections with the MAID analysis 
\cite{Maid} and with experimental data. For $\pi^{+}$
and $\pi^{0}$ photoproduction on the proton the data are taken 
from~\cite{Be+97,Leu00,Pre00} (MAMI), whereas for $\pi^{-}$
photoproduction we took the data from \cite{Fu+77} (Tokyo). 
In general, we obtain quite a good agreement with the data, especially 
in the region of the $\Delta$(1232) resonance at 330 MeV. 
Also in comparison with the MAID analysis our elementary production 
operator does quite well in this energy region. One notes 
only small discrepancies which very likely come from 
the fact that no other resonances besides the $\Delta$(1232) are included 
in our model. 

\begin{figure}[htb]
\includegraphics[scale=.7]{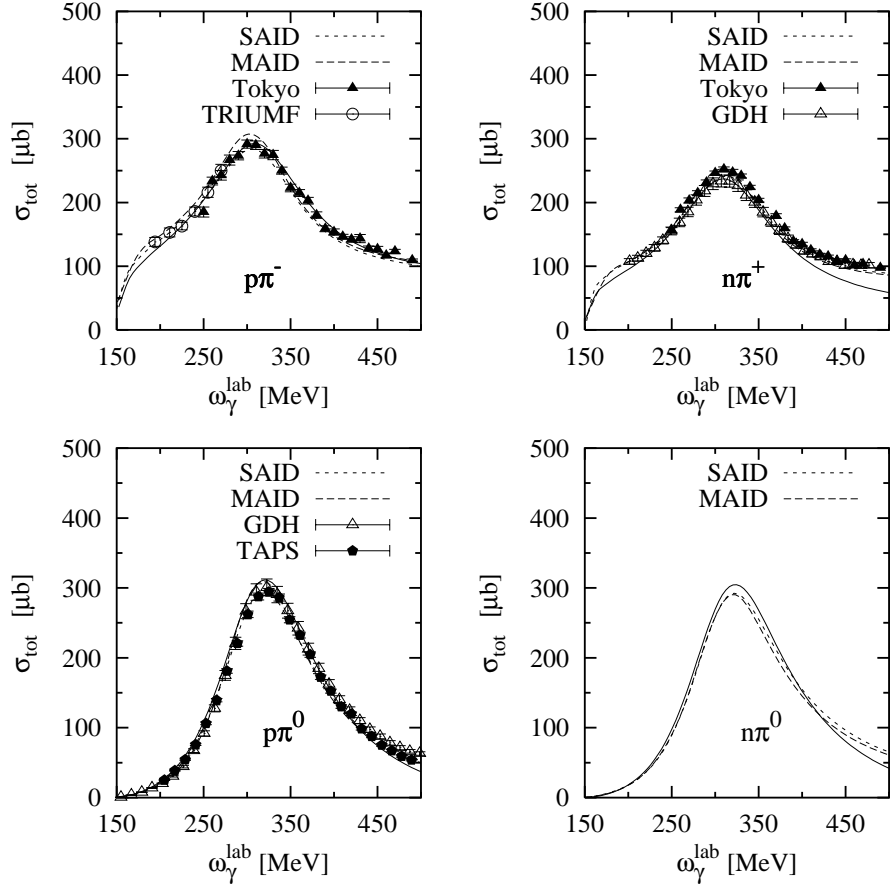}
\caption{Total cross sections for pion photoproduction on the nucleon
as a function of photon energy for all four physical
channels. Notation of the curves: solid: present model; short-dashed:
SAID~\protect\cite{Said}; dashed: MAID~\protect\cite{Maid}. Experimental 
data from \protect\cite{Fu+77} (Tokyo), \protect\cite{Bagheri88} (TRIUMF),
\protect\cite{Pre00} (GDH), \protect\cite{Hae96} (TAPS).} 
\label{tot}
\end{figure}

The total cross sections for the different pion channels are
shown in Fig.~\ref{tot} and compared with experimental data. In general, we
obtain a good agreement with the data using the small value
${f^2_{\pi N}}/{4\pi}=0.069$ for the pion-nucleon coupling constant. The
agreement with the data from \cite{Fu+77} and 
\cite{Bagheri88} for $\pi^-$ photoproduction on the neutron is again
satisfactory. In case of the $\pi^+$ photoproduction, the agreement 
is good up to a photon energy of about
400 MeV. For higher energies, the $D_{13}$ resonance, which is not
included in our calculation, gives a non-vanishing 
contribution~\cite{Maid}. The
$\pi^{+}$ data from \cite{Fu+77} are slightly underestimated in the
resonance region by our calculation but also by the MAID
analysis. Except for a tiny overestimation in the maximum, the
description of the data for $\pi^0$ production on the proton is also
very good. Therefore, this model for the elementary photoproduction
amplitude is quite satisfactory for our purpose, namely to incorporate
it into the reaction on the deuteron.

\section{Incoherent pion production on the deuteron}
\label{sec3}
We will briefly review the general formalism for incoherent pion
production on the deuteron. 
The general expression for the unpolarized differential cross
section of pion photoproduction reaction on the deuteron is given, using the
conventions of Bjorken and Drell \cite{BjD64}, by 
\begin{eqnarray}
\label{eq:3.2}
d\sigma = (2\pi)^{-5}\delta^{4}\left( k+d-p_{1}-p_{2}-q\right)
\frac{1}{|\vec{v}_{\gamma}-\vec{v}_{d}|} \frac{1}{2}
\frac{d^{3}q}{2\omega_{\vec{q}}} \frac{d^{3}p_{1}}{E_{1}}
\frac{d^{3}p_{2}}{E_{2}}
\frac{M_{N}^{2}}{4\omega_{\gamma}E_{d}}
\frac{1}{6}\sum_{s,m,t,m_{\gamma},m_d} 
|{\cal M}^{(t\mu)}_{s m\,m_{\gamma} m_d}|^{2} \, , 
\end{eqnarray}
where initial photon and deuteron four-momenta are denoted by
$k=(\omega_\gamma,\vec k\,)$ and $d=(E_d,\vec d\,)$, respectively, and
the four-momenta of final meson and two nucleons by $q=(\omega_q,\vec
q\,)$ with $\omega_{q} = \sqrt{m_{\pi}^{2} + \vec{q}^{\,2}}$, $m_\pi$
as pion mass, and
$p_j=(E_j,\vec p_j\,)$ ($j=1,2$) with $E_{j} =\sqrt{M_{N}^{2} +
\vec{p}_{j}^{\,2}}$, respectively, and $M_N$ as nucleon mass. Furthermore,  
$m_{\gamma}$ denotes the photon polarization, $m_{d}$ the spin 
projection of the deuteron, $s$ and $m$ total spin and projection
of the two outgoing nucleons, respectively, $t$ their total isospin,
$\mu$ the isospin projection of the pion, and $\vec{v}_{\gamma}$ and
$\vec{v}_{d}$ the velocities of photon and deuteron, respectively. 
The states of all particles are covariantly normalized.
The reaction amplitude is denoted by ${\cal M}^{(t\mu)}_{s\,m
m_{\gamma}m_d}$. As in~\cite{ScA96}, we have chosen as 
independent variables the pion
momentum $q$, its angles $\theta_{\pi}$ and $\phi_{\pi}$, the polar
angle $\theta_{p_{NN}}$ and the azimuthal angle $\phi_{p_{NN}}$ of the
relative momentum $\vec p_{NN}$ of the two outgoing nucleons as
independent variables. 

The total and relative momenta of the final $NN$-system are defined  
respectively by
\beq
\vec{P}_{NN} = \vec{p}_{1} + \vec{p}_{2}= \vec{k} - \vec{q}\,
\quad\mbox{and}\quad 
\vec p_{NN} = \frac{1}{2}\left(\vec{p}_{1} - \vec{p}_{2}\right)\, .
\label{eq:3.7}
\eeq
The absolute value of the relative momentum $\vec p_{NN}$ is given by 
\begin{eqnarray}
p_{NN} = \frac{1}{2}\sqrt{ \frac{ E_{NN}^{2}(W_{NN}^{2}-4
M_{N}^{2}) }{E_{NN}^{2}-P^{2}_{NN}\cos^{2}\theta_{Pp_{NN}}} }\, ,
\label{eq:3.11}
\end{eqnarray}
where $E_{NN}$ and $W_{NN}$ denote total energy and invariant mass of 
the $NN$ subsystem
\begin{eqnarray}
E_{NN} & = & E_{1}+E_{_2}=  \omega_{\gamma}+E_{d}-\omega_{q} \, , \nonumber \\
W_{NN}^{2} & = & {E}_{NN}^{2}-P^{2}_{NN}\,,
\label{eq:3.12}
\end{eqnarray}
and $\theta_{Pp_{NN}}$ is the angle between $\vec{P}_{NN}$ and $\vec p_{NN}$. 

\begin{figure}[htb]
\includegraphics[scale=1.0]{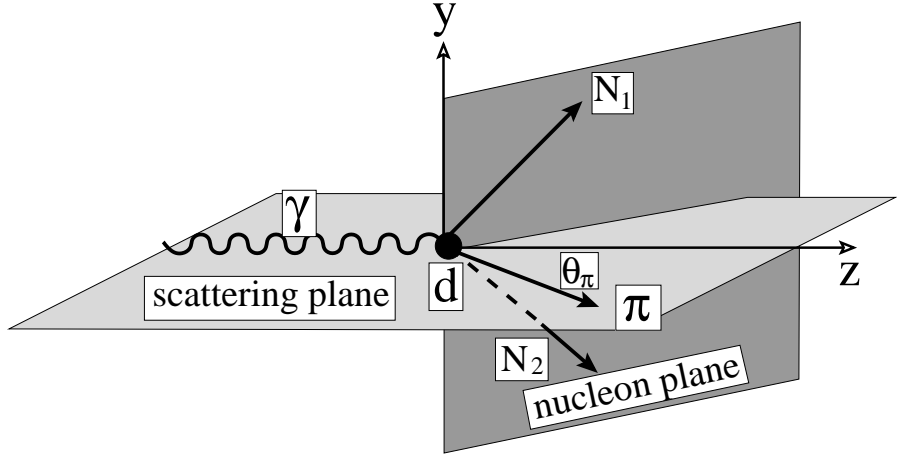}
  \caption{\small Kinematics in the laboratory system for  
    $\gamma d\rightarrow \pi NN$.}
  \label{labsys}
\end{figure}

For the evaluation we have chosen the laboratory frame where
$d^{\mu}=(M_d,\vec 0\,)$ with $M_d$ as deuteron mass. As coordinate
system a right-handed one is taken with $z$-axis along the
momentum $\vec k$ of the incoming photon and $y$-axis along 
$\vec k\times\vec q$. Thus the outgoing pion defines the
scattering plane. Another plane is defined by the momenta of the 
outgoing nucleons which we will call the nucleon plane 
(see Fig.~\ref{labsys}). 

In the later discussion of the main features of the processes we will
consider the semi-inclusive differential cross section $d^2\sigma
/d\Omega_{\pi}$, where only the final pion is detected. It is obtained
from the fully exclusive cross section  
\begin{eqnarray}
\frac{d^5\sigma}{d\Omega_{p_{NN}} d\Omega_{\pi} dq} = \frac{\rho_{s}}{6}
\sum_{s,m,t,m_{\gamma},m_{d}} |{\cal M}^{(t\mu)}_{s\,m\,m_{\gamma}m_d}|^{2}
\label{eq:3.13}
\end{eqnarray}
by integration over $q$ and $\Omega_{p_{NN}}$
\beq
\frac{d^2\sigma}{d\Omega_{\pi}}=\int_0^{q_{max}}dq
\int d\Omega_{p_{NN}}\,
\frac{d^5\sigma}{d\Omega_{p_{NN}} d\Omega_{\pi} dq}\,,
\eeq
where the maximal pion momentum $q_{max}$ is determined by the kinematics.
The phase space factor
$\rho_{s}$ in (\ref{eq:3.13}) is expressed in terms of relative and 
total momenta of the two final nucleons 
\begin{eqnarray}
\rho_{s} & = & \frac{1}{(2\pi)^{5}}\frac{p_{NN}^{2}M_{N}^{2}}
{\left| E_{2} (p_{NN}+\frac{1}{2} P_{NN} 
\cos\theta_{Pp_{NN}}) + E_{1}
(p_{NN}-\frac{1}{2} P_{NN} \cos\theta_{Pp_{NN}}) \right| }
\frac{q^{2}}{16\,\omega_{\gamma}M_{d}\,\omega_{q}} \, .
\label{eq:3.14}
\end{eqnarray}

\section{The transition matrix}
\label{sec4}

\begin{figure}[htb]
\includegraphics[scale=.7]{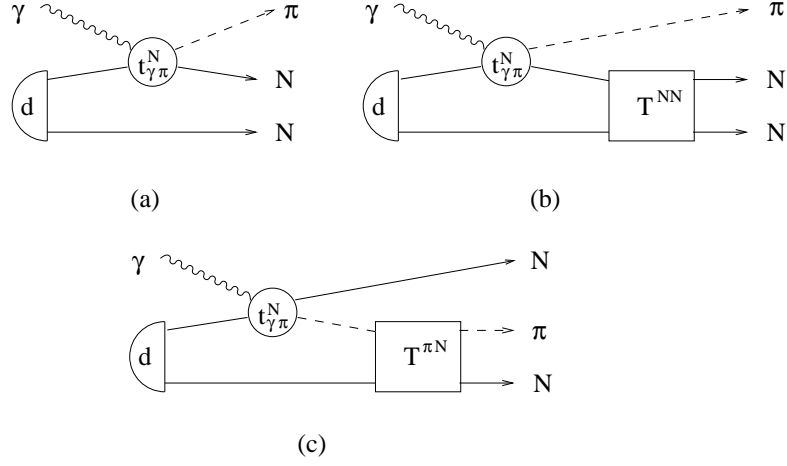}
\caption{Diagramatic representation of $\gamma d\rightarrow \pi NN$ 
including rescattering in the two-body subsystems of the final state: 
(a) impulse approximation (IA), (b) $NN$ rescattering, and (c) 
$\pi N$ rescattering.} 
\label{t-matrix}
\end{figure}

The general form of the photoproduction transition matrix is given by
\begin{eqnarray}\label{general}
{\cal M}^{(t\mu)}_{s\,m\,m_{\gamma}m_d}(\vec{k},\vec{q},\vec{p_1},\vec{p_2})
& = & ^{(-)}\bra{\vec{q}\,\mu,\vec{p_1}\vec{p_2}\,s\,m\,t-\mu}\epsilon_{\mu}
(m_{\gamma})J^{\mu}(0)\ket{\vec{d}\,m_d\,00}\, , 
\end{eqnarray}
where $J^{\mu}(0)$ denotes the current operator and
$\epsilon_{\mu}(m_{\gamma})$ the photon polarization vector. The
electromagnetic interaction consists of the elementary
production process on one of the nucleons $T_{\pi\gamma}^{(j)}$
$(j=1,2)$ and in principle a possible irreducible two-body production
operator $T_{\pi\gamma}^{(NN)}$. The final $\pi NN$ state is then
subject to the various hadronic two-body interactions as
described by an half-off-shell three-body scattering amplitude
$T^{\pi NN}$. In the following, we will neglect
the electromagnetic two-body production $T_{\pi\gamma}^{(NN)}$, and
the outgoing $\pi NN$ scattering state is approximated in this work by 
\beq
\ket{\vec{q}\,\mu,\vec{p_1}\vec{p_2}\,s\,m\,t-\mu}^{(-)}=
\ket{\vec{q}\,\mu,\vec{p_1}\vec{p_2}\,s\,m\,t-\mu} +G_{0}^{\pi NN (-)}
\,(T^{\pi N}(1)+T^{\pi N}(2)+T^{NN})
\ket{\vec{q}\,\mu,\vec{p_1}\vec{p_2}\,s\,m\,t-\mu}\,,
\eeq
where $\ket{\vec{q}\,\mu,\vec{p_1}\vec{p_2}\,s\,m\,t-\mu}$ denotes the 
free $\pi NN$ plane wave, $G_{0}^{\pi NN (-)}$ the free $\pi NN$
propagator, $T^{\pi N}(j)$ the reaction operator for  
$\pi N$-scattering on nucleon ``$j$'', and $T^{NN}$ the corresponding 
one for $NN$-scattering. 
This means, we include besides the pure impulse approximation (IA),
which is defined by the e.m.\ pion production on one of the nucleons alone,
only the complete rescattering by the final state interaction within each 
of the two-body subsystems. Therefore, the total transition matrix 
element reads in this approximation  
\begin{eqnarray}
\label{three}
{\cal M}^{(t\mu)}_{s\,m\,m_{\gamma}m_d} & = &
{\cal M}_{s\,m\,m_{\gamma}m_d}^{(t\mu)~IA} + 
{\cal M}_{s\,m\,m_{\gamma}m_d}^{(t\mu)~NN} + 
{\cal M}_{s\,m\,m_{\gamma}m_d}^{(t\mu)~\pi N}\, , 
\end{eqnarray}
in an obvious notation. A graphical representation of the transition
matrix is shown in Fig.~\ref{t-matrix}. We will now consider the different
contributions in detail. 

\subsection{The impulse approximation}
Here we briefly review the relevant formulae from~\cite{ScA96}. In
the IA the final state interaction is neglected and  
the pion and the NN-final states are described by pure plane
waves (see Fig.~\ref{t-matrix}(a)). For the spin $( |s m\rangle )$ 
and isospin $( |t -\mu\rangle )$ part of the two nucleon wave
functions we use a coupled spin-isospin basis $\ket{s m,\,t -\mu}$. 
The antisymmetric final NN plane wave function thus has the form
\begin{equation}
  |\vec{p}_{1},\vec{p}_{2},s m,t -\mu \rangle =
  \frac{1}{\sqrt{2}}(
    |\vec{p}_{1}\rangle^{(1)}|\vec{p}_{2}\rangle^{(2)} - (-)^{s+t}
    |\vec{p}_{2}\rangle^{(1)}|\vec{p}_{1}\rangle^{(2)})|s
  m\,,t -\mu\rangle\,,\label{plane_wave}
\end{equation}
where the superscript indicates to which particle the ket refers.
In the case of charged pions, only the $t = 1$ channel contributes 
whereas for $\pi^{0}$ production both $t = 0$ and $t = 1$ channels
have to be taken into account. Then the IA matrix element is given by 
\begin{eqnarray}
  {\cal M}_{s\,m\,m_{\gamma}m_d}^{(t\mu)~IA}
  (\vec{k},\vec{q},\vec{p_1},\vec{p_2})&=& \langle
\vec{p}_{1},\vec{p}_{2},s m,t -\mu 
  |t_{\gamma\pi}^{NN}(\vec k,\vec q\,)|\vec{d}m_{d},00 \rangle \nonumber\\
&=&\frac{1}{2} \int
  \frac{d^{3}p^{\prime}_{1}}{(2\pi)^{3}} \int
  \frac{d^{3}p^{\prime}_{2}}{(2\pi)^{3}}
  \frac{M_{N}^2}{E^{\,\prime}_{1}E^{\,\prime}_2}\nonumber\\
 && \sum_{m^{\prime}} \langle \,\vec{p}_{1}\vec{p}_{2},s
  m,t  -\mu |\, t_{\gamma\pi}^{NN}(\vec k,\vec q\,)
  |\,\vec{p}_{1}^{\,\prime}\vec{p}_{2}^{\,\prime}, 1 m^{\prime}, 0
  0 \rangle \langle\,\vec{p}_{1}^{\,\prime}\vec{p}_{2}^{\,\prime}, 1
  m^{\prime},00|\,\vec{d} m_{d},\,00\rangle \, . 
\end{eqnarray}
with 
\begin{equation}
t_{\gamma\pi}^{NN}(\vec k,\vec q\,)=t_{\gamma\pi}^{N(1)}(\vec k,\vec q\,)
+t_{\gamma\pi}^{N(2)}(\vec k,\vec q\,)\,,\label{t-IA}
\end{equation}
where $t_{\gamma\pi}^{N(j)}$ denotes the elementary production amplitude 
on nucleon ``$j$''. As mentioned above, we use covariant normalization for
the nucleon, deuteron and meson states, i.e.,
\begin{eqnarray}
\langle \vec{p}^{\,\prime} |\, \vec{p}\,\rangle & = &
(2\pi)^{3}\frac{E_{p}}{M_{N}}
\,\delta^{3}\!\left( \vec{p}^{\,\prime}-\vec{p}\,\right)\,,\quad
\langle \vec{d}^{\,\prime} |\, \vec{d}\,\rangle = (2\pi)^{3} 2E_{d}
\,\delta^{3}\,( \vec{d}^{\,\prime}-\vec{d}\,)\,,\quad
\langle \vec{q}^{\,\prime} |\, \vec{q}\,\rangle = (2\pi)^3\,2\,
\omega_q\,\delta(\vec q^{\,\prime}-\vec q)\,.
\end{eqnarray}
The deuteron wave function has the form
\begin{equation}
  \langle\,\vec{p}_{1}\vec{p}_{2}, 1
  m,\,00|\,\vec{d} m_{d},00\rangle = (2\pi)^{3}
  \delta^{3}(\,
    \vec{d}-\vec{p}_{1}-\vec{p}_2 \,)
  \frac{\sqrt{2\,E_{1}E_{2}}}
  {M_{N}} \widetilde{\Psi}_{m,m_{d}}(\vec{p}_{NN})
\end{equation}
with
\begin{equation}
  \widetilde{\Psi}_{m,m_{d}}(\vec{p}\,) =
  (2\pi)^{\frac{3}{2}}\sqrt{2E_{d}}
  \sum_{L=0,2}\sum_{m_{L}}i^{L}\,C^{L 1 1}_{m_{L} m m_{d}}\,
  u_{L}(p)Y_{Lm_{L}}(\hat{p}) \,,
\end{equation}
denoting with $C^{j_1 j_2 j}_{m_1 m_2 m}$ a Clebsch-Gordan coefficient. 
Using (\ref{t-IA}) one finds in the laboratory system for the IA-matrix
element the following expression
\begin{eqnarray}\label{g16}
  {\cal M}_{s\,m\,m_{\gamma}m_d}^{(t\mu)~IA}
  (\vec k,\vec q,\vec p_1,\vec p_2) &=&
 \sqrt{2}\sum_{m^{\prime}}\langle s 
  m,\,t -\mu|\,\Big( \langle
  \vec{p}_{1}|t_{\gamma\pi}^{N(1)}(\vec k,\vec q\,)|-\vec{p}_{2}\rangle
  \tilde{\Psi}_{m^{\prime},m_{d}}(\vec{p}_{2}) 
-(-)^{s+t}(\vec p_1 \leftrightarrow \vec p_2) 
\Big)\,|1
  m^{\prime},\,00\rangle \,.
\end{eqnarray}
Note that in (\ref{g16}) the elementary production operator acts on nucleon
``1'' only. This matrix element possesses the obvious symmetry under the
interchange of the nucleon momenta
\beq
  {\cal M}_{s\,m\,m_{\gamma}m_d}^{(t\mu)~IA}
  (\vec k,\vec q,\vec p_2,\vec p_1) =(-)^{s+t+1}\,
  {\cal M}_{s\,m\,m_{\gamma}m_d}^{(t\mu)~IA}
  (\vec k,\vec q,\vec p_1,\vec p_2) \,.\label{ia-symmetry}
\eeq

\subsection{$NN$ rescattering}

As next we will evaluate the $NN$-rescattering contribution whose 
Feynman diagram is shown in Fig.~\ref{t-matrix}~(b). The transition matrix 
element has the form 
\begin{eqnarray}
\label{tnn-fsi}
{\cal M}^{(t\mu)~NN}_{s\,m\,m_{\gamma}m_d} & = & 
\frac{1}{2} \int
  \frac{d^{3}p^{\prime}_{1}}{(2\pi)^{3}} \int
  \frac{d^{3}p^{\prime}_{2}}{(2\pi)^{3}}
  \frac{M_{N}^2}{E^{\,\prime}_{1}E^{\,\prime}_2}
\sum_{m^{\prime}}
{\cal R}^{NN,\, t \mu}_{s\,mm^{\prime}}(W_{NN},\vec p_1, \vec p_2,
\vec p^{\,\prime}_1, \vec p^{\,\prime}_2)\,\nonumber\\
& &  
{\cal G}_{0}^{\pi NN(+)}(E_{\gamma d},
\vec q,\vec p^{\,\prime}_1,\vec p^{\,\prime}_2)\,
{\cal M}_{s m^{\prime} m_{\gamma}m_d}^{(t\mu)~IA}(\vec k,\vec q,\vec
p^{\,\prime}_1,\vec p^{\,\prime}_2) 
\, .
\end{eqnarray}
Here ${\cal R}^{NN,\, t \mu}(W_{NN})$ contains the half-off-shell
$NN$-scattering  
matrix at the invariant energy of the $NN$-subsystem $W_{NN}$,
and ${\cal G}_{0}^{\pi NN(+)}(E_{\gamma d},\vec q,\vec p^{\,\prime}_1,
\vec p^{\,\prime}_2)$ denotes the free $\pi NN$ propagator. The latter
is given by 
\begin{eqnarray}
\label{pinn-propag}
{\cal G}_{0}^{\pi NN(+)}(E_{\gamma d},\vec{q},\vec{p}_1^{\,\prime},
\vec{p}_2^{\,\prime}) & = & \Big(E_{\gamma
    d}-\omega_{\pi}(\vec{q}\,)-E_{1}(\vec{p}_1^{\,\prime})-
    E_{2}(\vec{p}_2^{\,\prime}) + i\epsilon\Big)^{-1}\,,
\end{eqnarray}
where $E_{\gamma d}=\omega_\gamma +M_d$. Now we introduce relative 
and total momenta $\vec p_{NN}^{\,(\prime)}$ and $\vec P^{\,(\prime)}$, 
respectively, of the interacting nucleons in initial and final states
\beq
\vec p_{NN}^{\,(\prime)}=\frac{1}{2}\,(\vec{p}_1^{\,(\prime)}-
\vec{p}_2^{\,(\prime)})\,,\qquad 
\vec P^{\,(\prime)}=\vec{p}_1^{\,(\prime)}+
\vec{p}_2^{\,(\prime)}\,.
\eeq
Using nonrelativistic kinematics for the nucleons, one finds for 
the propagator
\begin{eqnarray}
\label{nonrelpropag}
{\cal G}_{0}^{\pi NN(+)}(E_{\gamma d},\vec{q},\vec{p}_1^{\,\prime},
\vec{p}_2^{\,\prime}) & = & 
\frac{M_N}{\widetilde p^{\,2} - {p_{NN}^{\prime 2}} + i\epsilon}\, ,
\end{eqnarray}
where $\widetilde p^{\,}$ is given by
\begin{eqnarray}
\widetilde p^{\,2} & = & M_N\Big( E_{\gamma
 d}-\omega_{\pi}(\vec{q}\,)-2M_N-\frac{(\vec{k}-\vec{q}\,)^2}{4M_N}\Big)
\,.
\end{eqnarray}
As next we separate the c.m.\ motion of the two-nucleon subsystem and 
obtain for the $NN$ rescattering amplitude ${\cal R}^{NN}$ 
\begin{eqnarray}
\label{tnn-hos1}
{\cal R}^{NN,\,t \mu}_{s m m^{\prime}}(W_{NN},\vec p_1, \vec p_2,
\vec p^{\,\prime}_1, \vec p^{\,\prime}_2)
& = & 2(2\pi)^6 \delta^3(\vec P^{\,\prime}-\vec P)
\frac{\sqrt{E_1 E_2 E_1' E_2'}}{M_N^2}\, 
\widetilde {\cal R}_{s m m^{\prime}}^{NN,\,t\mu}
(W_{NN},\vec p_{NN},\vec p^{\,\prime}_{NN})\,.
\end{eqnarray}
Here we have introduced the conventional $NN$-scattering matrix
$\widetilde {\cal R}_{s\,mm^{\prime}}^{NN,\,t\mu}$ with respect to
noncovariantly normalized states, which is expanded in terms of the
partial wave contributions ${\cal
T}_{Js\ell\ell^{\prime}}^{NN,\,t\mu}$ 
\begin{eqnarray}
\label{tnn-hos}
\widetilde {\cal R}_{s\,mm^{\prime}}^{NN,\,t\mu}(W_{NN},\vec p_{NN},\vec
 p^{\,\prime}_{NN}) 
 & = & \sum_{J\ell\ell^{\prime}}
 {\cal F}_{\ell\ell^{\prime}\,mm'}^{NN,\,Js}
 (\hat{p}_{NN},\hat{p}_{NN}^{\,\prime}) 
{\cal T}_{Js\ell\ell^{\prime}}^{NN,\,t\mu}
 (W_{NN},p_{NN},p_{NN}^{\,\prime})\, , 
\end{eqnarray}
where orbital and total angular momenta of the two-nucleon system are 
denoted by $\ell$ and $J$, respectively. The purely angular function ${\cal
F}_{\ell\ell^{\prime} mm^{\prime}}^{NN,\,Js}
(\hat{p}_{NN},\hat{p}_{NN}^{\,\prime})$ is defined by  
\begin{eqnarray}
\label{tnn-ff}
{\cal F}_{\ell\ell^{\prime} mm^{\prime}}^{NN,\,Js}(\hat{p}_{NN},
\hat{p}_{NN}^{\,\prime}) & = & 
 \sum_{M m_{\ell}m_{\ell^{\prime}}} C^{\ell s J}_{m_{\ell} m M}\,
C^{\ell^{\prime} s J}_{m_{\ell^{\prime}} m^{\prime} M}
 Y^{\star}_{\ell m_{\ell}}(\hat{p}_{NN})
 Y_{\ell^{\prime} m_{\ell^{\prime}}}(\hat{p}_{NN}^{\,\prime})\,.
\end{eqnarray}
Collecting the various pieces and substituting
(\ref{tnn-hos1}) and 
(\ref{nonrelpropag}) into (\ref{tnn-fsi}), one obtains the following 
expression for the $NN$ rescattering contribution
\begin{eqnarray}
\label{tnn-fsi-final}
{\cal M}^{(t\mu)~NN}_{s\,m\,m_{\gamma}m_d}
(\vec k,\vec q,\vec p_1,\vec p_2) & = & 
\sum_{m^{\prime}}\int d^3\vec p^{\,\prime}_{NN} 
\sqrt{\frac{E_1 E_2}{E_1' E_2'}}
\,\widetilde {\cal R}_{s m m^{\prime}}^{NN,\,t\mu}(W_{NN},\vec p_{NN},\vec
p^{\,\prime}_{NN}) \nonumber\\
&&\frac{M_N}{\widetilde p^{\, 2} - p_{NN}^{\prime\,2} + i\epsilon}
{\cal M}^{(t\mu)~IA}_{s\,m^{\prime},m_{\gamma}m_d}(\vec k,\vec
q,\vec p^{\,\prime}_1,\vec p^{\,\prime}_2)\,. 
\end{eqnarray}
where $\vec p^{\,\prime}_{1/2}=\pm \vec p^{\,\prime}_{NN} + (\vec
k-\vec q\,)/2$  and $E_{1/2}'$ the corresponding on-shell energies.

\subsection{$\pi N$ rescattering}

The last contribution concerns the $\pi N$-rescattering in the final state 
whose diagram is shown in Fig.~\ref{t-matrix}~(c). The corresponding
transition matrix element has formally a similar structure as the one for 
$NN$-rescattering and is given by
\begin{eqnarray}
\label{tpn-fsi}
M^{(t\mu)~\pi N}_{s\,m,m_{\gamma}m_d} 
(\vec k,\vec q,\vec p_1,\vec p_2) & = &\frac{1}{2}\,
\sum_{\alpha'} 
\int \frac{d^3\vec q^{\,\prime}}{(2\pi)^3} 
\frac{d^3\vec p^{\,\prime}_1}{(2\pi)^3} 
\frac{d^3\vec p^{\,\prime}_2}{(2\pi)^3} 
\frac{M^2_N}{2\omega_{q^{\prime}}E_1' E_2'}\nonumber\\
&&\Big[{\cal R}^{\pi N}_{\alpha \alpha'}(\vec{q},\vec{p}_1,\vec{p}_2,
\vec q^{\,\prime},\vec p^{\,\prime}_1,\vec p^{\,\prime}_2)\,
-(-)^{s+t}{\cal R}^{\pi N}_{\alpha \alpha'}
(\vec{q},\vec{p}_2,\vec{p}_1,
\vec q^{\,\prime},\vec p_1^{\,\prime},\vec p_2^{\,\prime})\,
\Big]\nonumber\\
&& 
{\cal G}_{0}^{\pi NN(+)}(E_{\gamma d},
\vec q^{\,\prime},\vec p_1^{\,\prime},\vec p_2^{\,\prime})\,
{\cal M}_{s'm^{\prime} m_{\gamma}m_d}^{(t'\mu')\,IA}(\vec k,
\vec q^{\,\prime},\vec p_1^{\,\prime},\vec p_2^{\,\prime}) \, ,
\end{eqnarray}
where we have introduced as a shorthand for the quantum numbers
$\alpha=(s\,m t\mu)$ and have made use of the symmetry (\ref{ia-symmetry}). 
Furthermore,
${\cal R}^{\pi N}_{\alpha \alpha'}(\vec{q},\vec{p}_1,\vec{p}_2,
\vec q^{\,\prime},\vec p^{\,\prime}_1,\vec p^{\,\prime}_2)$
containes the half-off-shell $\pi N$-scattering matrix.
Separating the non-participating spectator nucleon and the c.m.\ motion 
of the interacting $\pi N$ subsystem, switching to an uncoupled 
spin-isospin basis, and coupling the isospins of the interacting pion 
and nucleon to a total isospin $\tilde{t}$, one obtains 
\beqa
\label{tpn-hos1}
{\cal R}^{\pi N}_{\alpha \alpha'}
(\vec{q},\vec{p_1},\vec{p_2},\vec q^{\,\prime},\vec p^{\,\prime}_1,
\vec p^{\,\prime}_2)&=&
(2\pi)^9\,2\,\sqrt{\omega_q\omega_{q'}}\,\frac{E_1}{M_N}\,
\sqrt{\frac{E_2 E_2'}{M_N^2}}\,\delta(\vec{p}_1-\vec p^{\,\prime}_1)
\,\delta(\vec q+\vec{p}_2-\vec q^{\,\prime}-\vec p^{\,\prime}_2)
\nonumber\\
&&\sum_{m_2m_2'}\sum_{\mu_2\mu_2'}\sum_{\tilde t\tilde \mu}\,
{\cal C}^{\tilde t\tilde\mu}_{\alpha \alpha'}(m_2,m_2',\mu_2,\mu_2')\,
\widetilde {\cal R}_{m_2m_2'}^{\pi N,\,\tilde t\tilde \mu}
(W_{\pi N}(\vec p_2),\vec p_{\pi N},\vec  p^{\,\prime}_{\pi N}) \,,
\eeqa
where 
\beq
{\cal C}^{\tilde t\tilde\mu}_{\alpha \alpha'}(m_2,m_2',\mu_2,\mu_2')=
C^{1\frac{1}{2}\tilde t}_{\mu \mu_{2} \tilde\mu}\,
C^{1\frac{1}{2}\tilde t}_{\mu' \mu_{2}' \tilde\mu}\,
 \sum_{m_1 }
C^{\frac{1}{2}\frac{1}{2}s}_{m_1 m_{2} m}\,
C^{\frac{1}{2}\frac{1}{2}s}_{m_1 m_{2}' m'}\,
\sum_{\mu_1 }
C^{\frac{1}{2}\frac{1}{2}t}_{\mu_1 \mu_{2} -\mu}\,
C^{\frac{1}{2}\frac{1}{2}t'}_{\mu_1 \mu_{2}' -\mu'}\,
\eeq
contains the recoupling coefficients, and 
$\widetilde {\cal R}^{\pi N,\,\tilde t \tilde\mu}_{m_2m_2'}$ 
denotes the half-off-shell $\pi N$-scattering matrix at the 
invariant mass $W_{\pi N}(\vec p_2)=
\sqrt{(E_2+\omega_q)^2-(\vec q+\vec p_2)^2}$ of the $\pi N$ subsystem. 
Furthermore, $m_{2}$ ($m'_{2}$) and $\mu_{2}$ ($\mu'_{2}$)
denote the spin and isospin projections of the final (initial) nucleon 
in the $\pi N$ subsystem, respectively. The relative momentum of the 
final (initial) pion-nucleon subsystem is given, respectively, by
\beq
\label{pinprelative}
\vec p_{\pi N} = 
\frac{M_N\vec q-m_{\pi}\vec p_2}{M_N+m_{\pi}}\,,\qquad 
\vec  p^{\,\prime}_{\pi N}=
\frac{M_N\vec q^{\,\prime}-m_{\pi}\vec p_2^{\,\prime}}{M_N+m_{\pi}}=
\frac{M_N}{M_N+m_{\pi}}\,(\vec q+\vec p_2)-\vec p_2^{\,\prime}
\,.
\eeq
The $\pi N$ scattering matrix is now expanded in terms of partial 
wave amplitudes
\beqa\label{tpin_partial}
\widetilde {\cal R}^{\pi N,\,\tilde t \tilde\mu}_{m_2m_2'}
(W_{\pi N}(\vec p_2),\vec p_{\pi N},\vec p^{\,\prime}_{\pi N})&=&
\sum_{J\ell}
{\cal F}_{J\ell m_2m_2'}^{\pi N} 
(\hat p_{\pi N},\hat p^{\,\prime}_{\pi N})
{\cal T}^{\pi N,\,\tilde t \tilde\mu}_{J\ell}
(W_{\pi N}(\vec p_2), p_{\pi N}, p^{\,\prime}_{\pi N})\,,
\eeqa
where we have defined
\beqa
{\cal F}_{J\ell m_2m_2'}^{\pi N} 
(\hat p_{\pi N},\hat p^{\,\prime}_{\pi N})
& = & 
\sum_{m_\ell m_\ell' M} C^{\frac{1}{2} \ell J}_{ m_2 m_\ell M}\,
C^{\frac{1}{2} \ell J}_{ m_2' m_\ell' M}\,
Y_{\ell m_{\ell}}^{\star}(\hat{{p}}_{\pi N})Y_{\ell m_{\ell}'}(\hat{p}
^{\prime}_{\pi N})\,.
\eeqa
Inserting (\ref{tpn-hos1}) with (\ref{tpin_partial}) into 
(\ref{tpn-fsi}), one obtains the
final form for the $\pi N$-rescattering contribution 
\begin{eqnarray}
\label{tpn-hos21}
M^{(t\mu)~\pi N}_{s\,m\,m_{\gamma}m_d}(\vec k,\vec q,\vec p_1,\vec p_2) & = &
\frac{1}{2}\,\sum_{\alpha'} \int
d^3\vec{p}_2^{\,\prime}\sqrt{\frac{\omega_q\,E_2}{\omega_{q'}\,E_2'}}
\sum_{m_2 m_2'}\sum_{\mu_2 \mu_2'}\sum_{\tilde t\tilde\mu}
{\cal C}^{\tilde t\tilde\mu}_{\alpha \alpha'}(m_2,m_2',\mu_2,\mu_2')
\nonumber\\ &&
\Big[\sum_{J\ell}
{\cal F}_{J\ell m_2m_2'}^{\pi N} 
(\hat p_{\pi N},\hat p^{\,\prime}_{\pi N})
{\cal T}^{\pi N,\,\tilde t \tilde\mu}_{J\ell}
(W_{\pi N}(\vec p_2), p_{\pi N}, p^{\,\prime}_{\pi N})\,
{\cal G}_{0}^{\pi NN(+)}
(E_{\gamma d},\vec q^{\,\prime},\vec p_1,\vec
p^{\,\prime}_2)\,
\nonumber\\ &&
{\cal M}^{(t^{\prime}\mu^{\prime})~IA}
_{s^{\prime}m^{\prime},m_{\gamma}m_d}
(\vec k,\vec q^{\,\prime},\vec p_1,\vec p^{\,\prime}_2)
-(-)^{s+t}(\vec p_1 \leftrightarrow \vec p_2)\Big]\, , 
\end{eqnarray}
where ${\vec{p}^{\,\prime}_{\pi N}}$ and ${\vec{p}_{\pi N}}$ 
are given in (\ref{pinprelative}) and 
$\vec q^{\,\prime}=\vec q+\vec p_2-\vec p^{\,\prime}_2$.

\section{Results and discussion}
\label{sec5}
The three contributions to the pion production amplitude, i.e., the IA
in (\ref{g16}) and the two rescattering contributions in
(\ref{tnn-fsi-final}) and (\ref{tpn-hos21}) are evaluated by taking a
realistic $NN$ potential model for the deuteron wave function and the
$NN$ scattering amplitudes, in this work the Paris potential. 
Specifically, we have taken the deuteron wave function from~\cite{La+81}
and the interaction in the separable representation of~\cite{HaP84,HaP85}. 
Explicitly, we have included all partial waves with
total angular momentum $J\le 3$. Also in the case of $\pi N$
rescattering we have used the separable energy-dependent $\pi N$
potential of~\cite{NoB90} and have considered all $S$-, $P$- and
$D$-waves. The remaining three dimensional integrals in
(\ref{tnn-fsi-final}) over $\vec p^{\,\prime}_{NN}$ and in
(\ref{tpn-hos21}) over $\vec{p}_2^{\,\prime}$ are evaluated
numerically. We would like to remark, that we have obtained essentially
the same results if we take the Bonn r-space potential~\cite{MaH87} 
instead of the Paris one.

The discussion of our results is divided into two parts. First, 
we will discuss the influence of FSI on the total cross section by
comparing the pure IA with the inclusion of two-body rescattering in
the final state. Furthermore, we will confront our results with  
experimental data and other theoretical calculations. In the second
part, we will then consider the semi-exclusive differential cross section
$d^2\sigma /d\Omega_{\pi}$ where only the pion is detected in the
final state. 

\subsection{Total Cross Section}
\begin{figure}[htb]
\includegraphics[scale=.7]{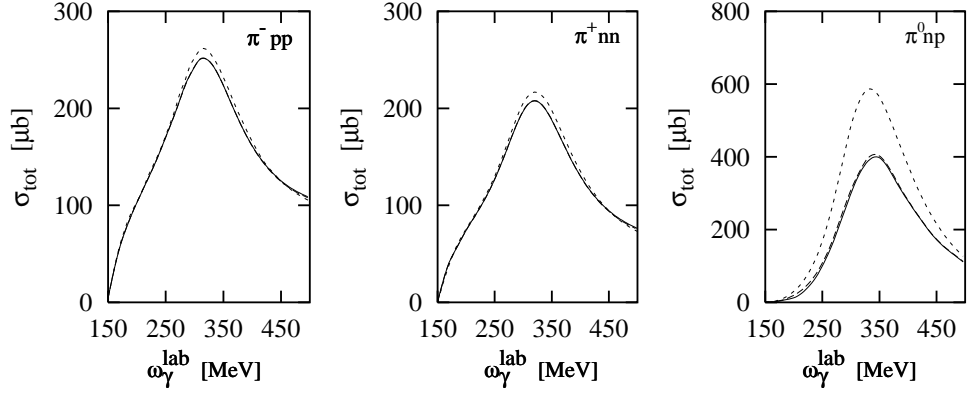}
\caption{Total cross sections for $\gamma d\rightarrow\pi NN$.
Notation of curves: short-dashed: impulse approximation (IA); dashed: 
IA plus $NN$ rescattering; solid: IA plus $NN$ and $\pi N$
rescattering. The left, middle and right panels represent the 
results for $\gamma d\rightarrow \pi^-pp$, $\pi^+nn$ and $\pi^0np$,
respectively.} 
\label{tot_our}
\end{figure}

\begin{figure}[htb]
\includegraphics[scale=.7]{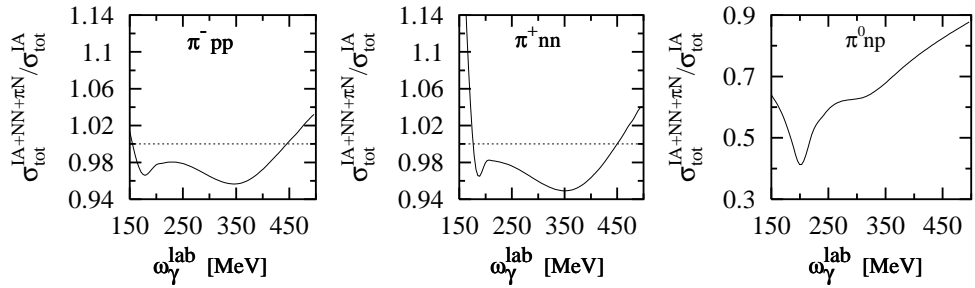}
\caption{\small The ratio of the total cross section with complete
rescattering $\sigma_{tot}^{IA+NN+\pi N}$ to the one in the IA
$\sigma_{tot}^{IA}$ as a function of the lab photon energy. The left,
middle and right panels represent the ratios for $\pi^-$, $\pi^+$ and
$\pi^0$ production, respectively.} 
\label{tot_ratio_our_r1}
\end{figure}

\begin{figure}[htb]
\includegraphics[scale=.7]{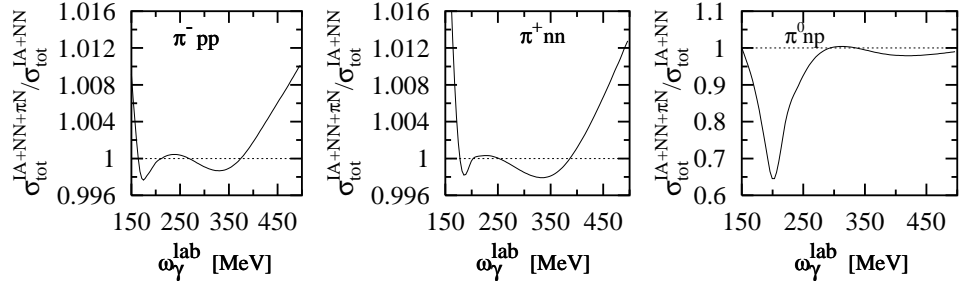}
\caption{\small The ratio  of the total cross section with complete
rescattering $\sigma_{tot}^{IA+NN+\pi N}$ to the one, where only 
$NN$-rescattering is included, $\sigma_{tot}^{IA+NN}$ as 
function of the lab photon energy. The left, middle and right panels
represent the ratios for $\pi^-$, $\pi^+$ and $\pi^0$ production,
respectively.}  
\label{tot_ratio_our_r2}
\end{figure}

Our results for the total cross sections in IA alone and with FSI
effects included are presented in Fig.~\ref{tot_our}. 
In order to show in greater detail the relative influence of
rescattering effects on the total cross sections, we show in
Fig.~\ref{tot_ratio_our_r1} the effect of complete rescattering
relative to the IA by the ratio $\sigma_{tot}^{IA+NN+\pi
N}/\sigma_{tot}^{IA}$ and in Fig.~\ref{tot_ratio_our_r2} the effect of
$\pi N$-rescattering alone relative to the complete effect by the ratio
$\sigma_{tot}^{IA+NN+\pi N}/\sigma_{tot}^{IA+NN}$, where
$\sigma_{tot}^{IA}$ denotes the total cross section in the impulse
approximation, $\sigma_{tot}^{IA+NN}$ the one including only $NN$
rescattering, and $\sigma_{tot}^{IA+NN+\pi N}$ the one including
in addition $\pi N$ rescattering contributions. One readily notes the
importance of rescattering effects, in particular for the 
$\pi^0$ channel. FSI leads in all cases, to a reduction of the total 
cross section, except close to the production threshold, where for
charged pions one notes a sizeable increase and above about 450
MeV. The sizeable effect of $\pi N$ rescattering in the threshold region
confirms the previous results in~\cite{BlL77} for the coherent 
reaction and in~\cite{KoW77,Le+00} for the incoherent one. Furthermore, it
has been pointed out already in~\cite{Le+00}, that also for charged
pion channels rescattering effects are important in the threshold
region.

In the energy range of the 
$\Delta$(1232) resonance, one finds the strongest reduction by
rescattering effects which arise predominantly from $NN$
rescattering, whereas the influence of $\pi N$ rescattering appears
almost negligible, about an order of magnitude smaller, as is evident
from Fig.~\ref{tot_ratio_our_r2}. Only for neutral pion production
$\pi N$ rescattering becomes noticable below the $\Delta$ region  
and amounts to about 40 percent of the total effect near 175 MeV. The
reason for this relatively small effect from $\pi N$ rescattering lies
in the much smaller $\pi N$-interaction 
in comparison to the $NN$-interaction. This is manifest by the fact that
the $S$-wave scattering length of $\pi N$-scattering is
about two orders of magnitude smaller than the one of
$NN$-scattering. 

While for charged pion photoproduction FSI effects are 
relatively small, not more than about 5 percent, they are quite strong
in the case of neutral pion photoproduction, reaching a maximum of
about 60 percent at 175 MeV and still about 35 percent in the $\Delta$
region. The large $NN$ FSI effect in neutral pion production has two
related sources. The first arises from the fact that for the $\pi^0$
channel the IA contains a contribution from the coherent process
because the final $NN$ plane wave contains a deuteron bound state
component. This part, which is absent for charged pion production, is
automatically excluded as soon as the $NN$ 
interaction is switched on, because the scattering state is orthogonal
to the deuteron ground state. The second source is the change in the
radial wave function of the final $NN$ partial waves by the
interaction. The latter is also responsible for the reduction of the
charged pion channels. 

\begin{figure}[htb]
\includegraphics[scale=.7]{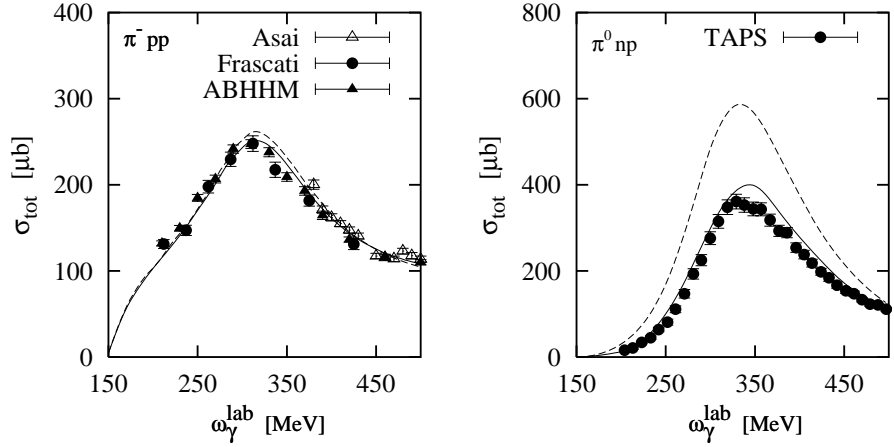}
\caption{Total cross sections for $\pi^-$ (left panel) and
$\pi^0$ (right panel) photoproduction on the deuteron. Solid
curves: our results with $NN$ and $\pi N$ rescattering; dashed curves: 
impulse approximation. Experimental data from
\protect\cite{Be+73} (ABHHM), \protect\cite{ChD75} (Frascati)
and \protect\cite{As+90} (Asai) for $\pi^-$, and from
\protect\cite{Kru99} (TAPS) for $\pi^0$ production.} 
\label{tot_deut_exp}
\end{figure}

Fig.~\ref{tot_deut_exp} shows a comparison of our results for the
total cross sections for $\pi^-$ and $\pi^0$ photoproduction with
experimental data. In view of the fact that data for $\pi^{+}$
production in the $\Delta$ region are not available, we
concentrate the discussion on $\pi^{-}$ and $\pi^0$ data. In the case
of $\pi^-$ production we have taken the experimental data
from~\cite{Be+73,ChD75,As+90} while for $\pi^0$ production we compare
our results with the experimental data from~\cite{Kru99}. 
One readily notes, that in agreement with earlier results the pure IA
cannot describe the experimental data, especially in the case of
$\pi^0$ production. The inclusion of such effects improves the
agreement between experimental data and theoretical predictions
considerably. Only in the maximum of $\pi^0$ production our model
overestimates the measured total cross section by about
6$\%$. 

\begin{figure}[htb]
\includegraphics[scale=.7]{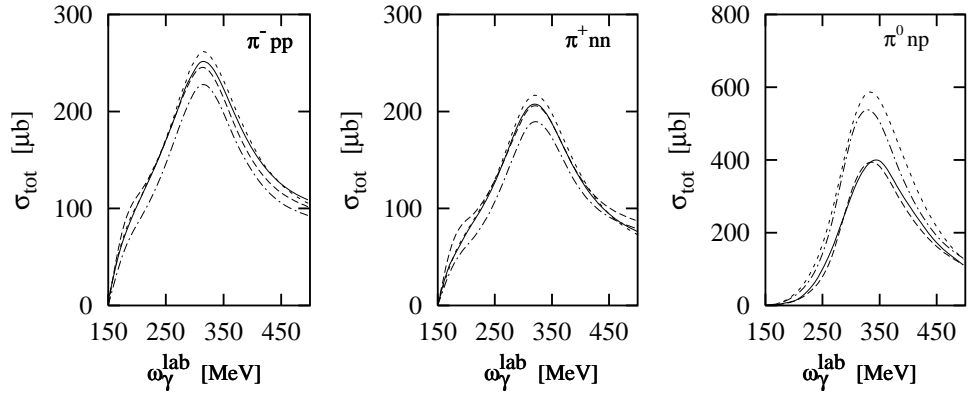}
\caption{Total cross sections for pion photoproduction on the deuteron.
Short-dashed curves: IA of present calculation; solid curves: IA plus $NN$
and $\pi N$ rescattering of present calculation; dash-dotted curves:
IA of Levchuk {\it et al.}~\protect\cite{LeS00}; dashed curves: IA
plus $NN$ rescattering from~\protect\cite{LeS00}.} 
\label{tot_deut_theory}
\end{figure}

Finally, we compare our results with the theoretical predictions
from~\cite{LeS00} as shown in Fig.~\ref{tot_deut_theory}.  
Surprisingly, one notes for the impulse approximation a significant 
difference for charged pion production which cannot be attributed to 
the use of different elementary production operators. A possible 
explanation for this feature will be presented in the next section, 
where we will discuss the differential cross sections. On the other hand, 
inclusion of FSI leads in our case to a reduction whereas an enhancement
was found in~\cite{LeS00} for charged pion production. As a result, one 
finds quite close agreement for the two calculations, if FSI is included.
On the contrary, for neutral pion production we obtain reasonable 
agreement in the IA with~\cite{LeS00}, only in the maximum and 
at higher energies one notes some differences which very likely come from 
the neglect of higher resonances in our elementary photoproduction 
model. Both calculations 
predict a strong reduction by FSI leading to a satisfactory agreement.
The remaining small differences probably stem from different pion 
photoproduction operators and from different realistic $NN$ potential 
models, since in \cite{LeS00} the Bonn r-space potential
model~\cite{MaH87} has been used. 

\subsection{Differential Cross Section}\label{chap:4:2}

\begin{figure}[htb]
\includegraphics[scale=.6]{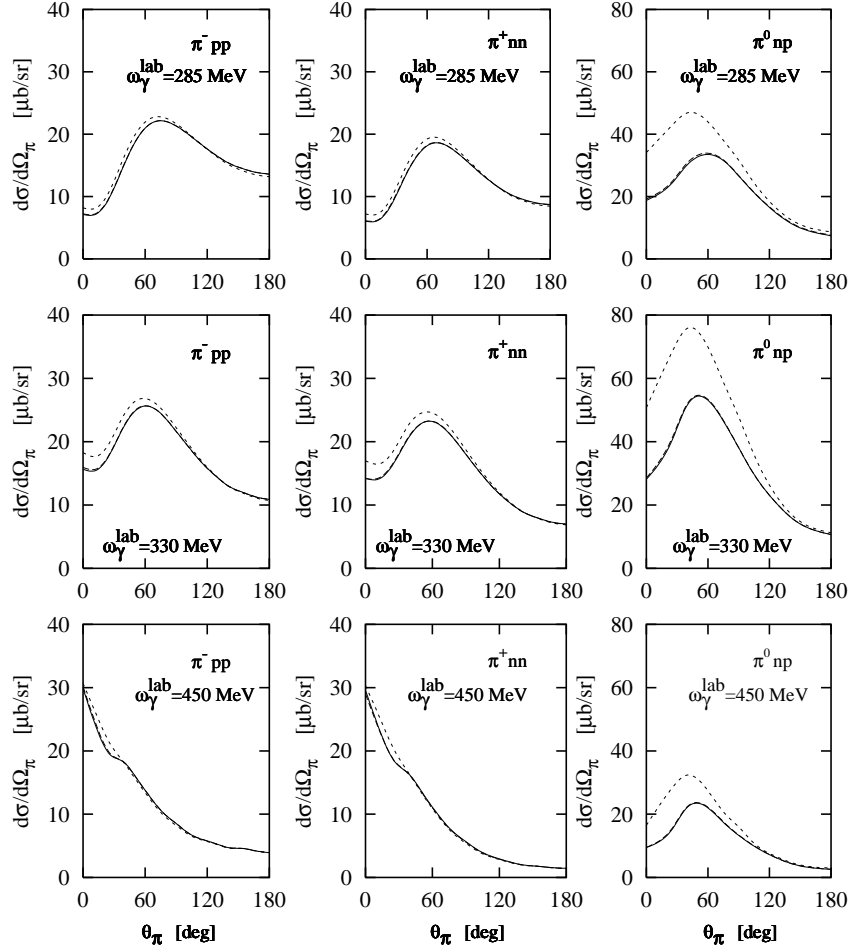}
\caption{Differential cross sections for pion photoproduction on the
deuteron for various energies. Short-dashed curves: IA; dashed curves: IA plus $NN$
rescattering; solid curves: IA plus $NN$ and $\pi N$ rescattering. The
left, middle and right panels represent the differential cross section
for $\gamma d\rightarrow \pi^-pp$, $\pi^+nn$ and $\pi^0np$,
respectively.}   
\label{diffour} 
\end{figure}

We begin the discussion by presenting the results for the 
differential cross sections in the pure IA and with rescattering included 
in Fig.~\ref{diffour}. Here one sees that the major contribution from FSI
appears at forward pion angles, i.e., at angles less than 90$^\circ$, 
predominantly from $NN$ rescattering,
whereas rescattering effects become quite small for backward angles. 
As already noted for the total cross section, the overall effect is quite 
small for charged pions, reaching a maximal reduction at $\theta_\pi=0^\circ$
of about 15 percent and decreasing rapidly with increasing angle. 
In the case of the $\pi^0$ channel, the results for the total cross
section have already shown that the effect of rescattering is
large. Again one notes that the dominant effect appears at 
forward pion angles resulting in a strong reduction of the order of 40
percent. At 90$^\circ$ the reduction is still sizeable but decreases
to a tiny effect at 180$^\circ$. 

\begin{figure}[htb]
\includegraphics[scale=.7]{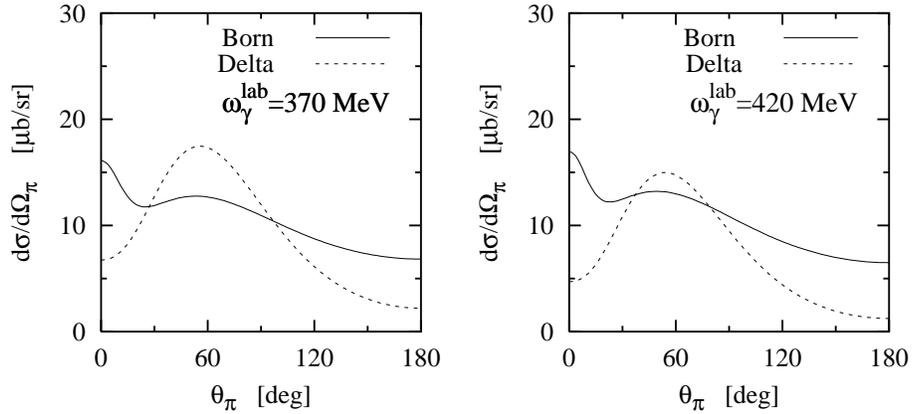}
\caption{Differential cross section for $\pi^-$
photoproduction on the deuteron in impulse approximation. Solid
curves: contribution of Born terms alone; short-dashed curves: contribution
of the $\Delta$(1232) resonance alone.} 
\label{borndelta}
\end{figure}

Another interesting feature is that for charged pion production 
in contrast to $\pi^0$ production 
the angular distribution of the emitted pion is more and more forward 
peaked with increasing photon energy. Its origin are the Born terms
which are absent for $\pi^0$ production. 
This is demonstrated in Fig.~\ref{borndelta}
where the separate contributions from Born and resonance terms are shown. 
More than 70 percent of the differential cross section
at $\theta_{\pi}=0$ comes from the Born terms. 

\begin{figure}[htb]
\includegraphics[scale=.7]{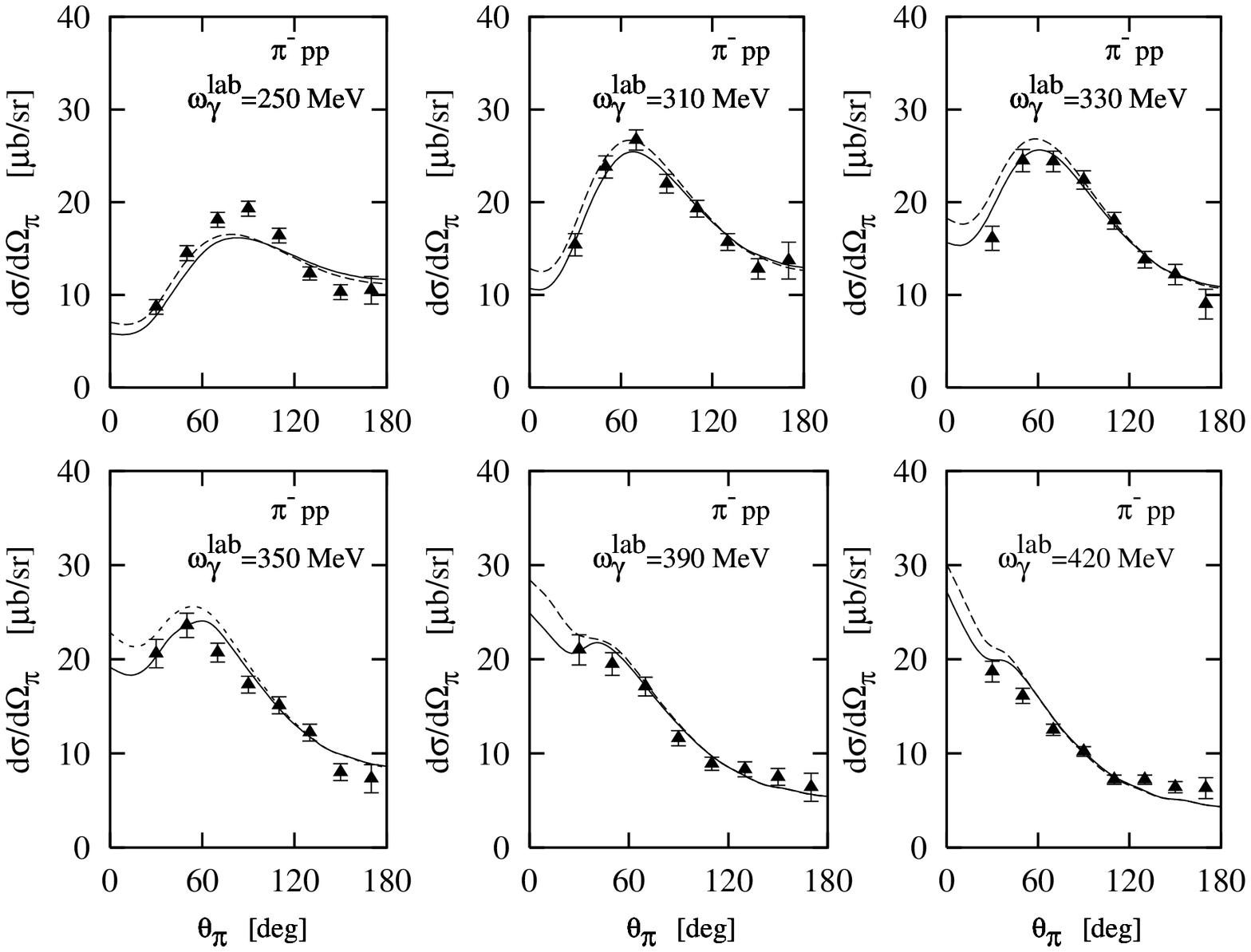}
\caption{Differential cross sections for $\pi^-$  photoproduction on the
deuteron. Solid curves: IA plus $NN$ and
$\pi N$ rescattering; dashed curves: IA. Experimental data: 
\protect\cite{Be+73}.}  
\label{diffwithexp-}
\end{figure}

\begin{figure}[htb]
\includegraphics[scale=.7]{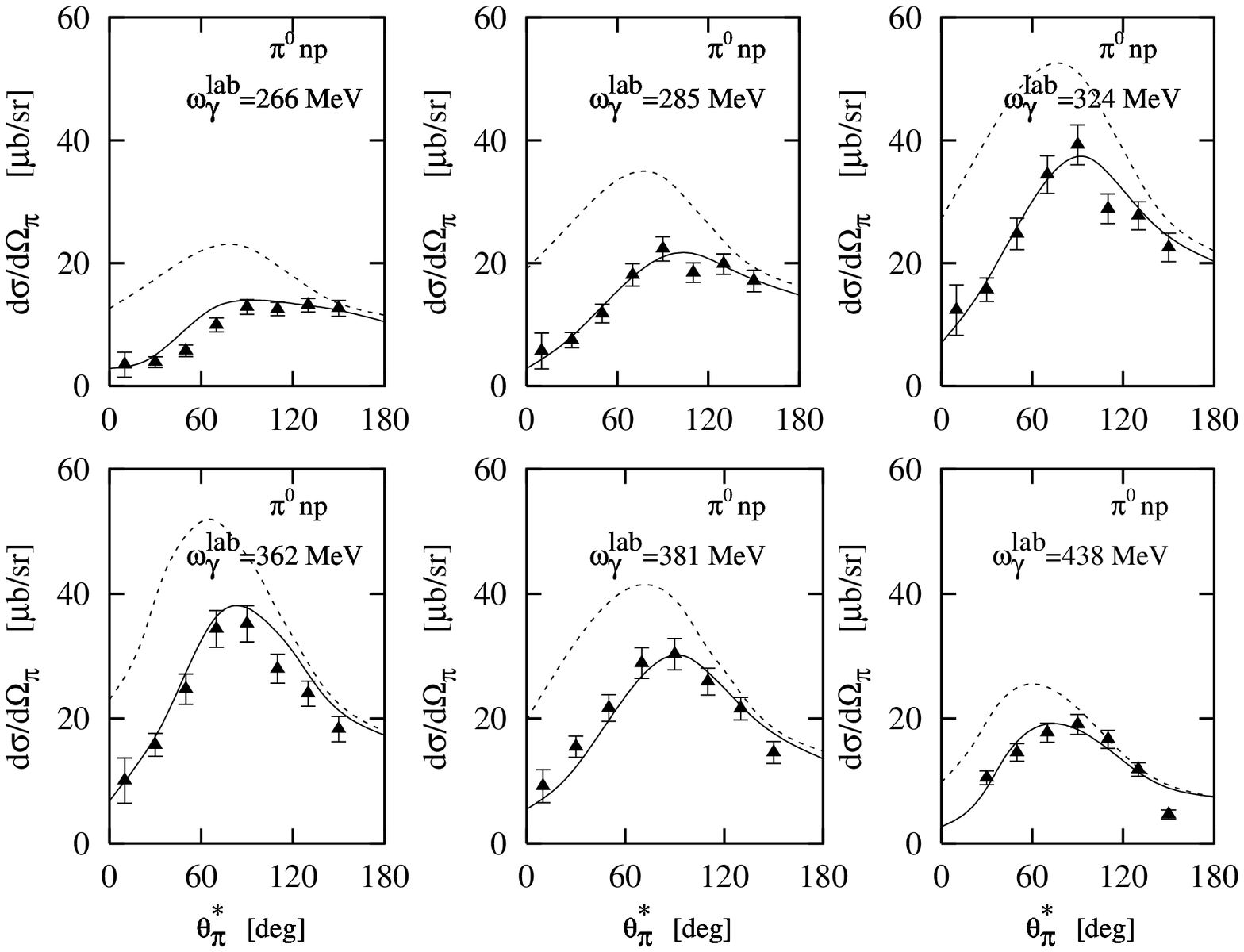}
\caption{Differential cross sections for $\pi^0$ photoproduction on the
deuteron. Solid curves: IA plus $NN$ and
$\pi N$ rescattering; dashed curves: IA. Experimental data: 
\protect\cite{Kru99}.}  
\label{diffwithexp+}
\end{figure}

A comparison with experimental data from \cite{Be+73} for $\pi^-$ production 
and from \cite{Kru99} for $\pi^0$ production is shown in 
Figs.~\ref{diffwithexp-} and \ref{diffwithexp+}. 
Since in~\cite{Kru99} the differential cross sections for the reaction
$d(\gamma,\pi^0)np$ are given in the so-called $\gamma N$ c.m.\
frame we have transformed the differential cross sections
from the lab frame to the $\gamma N$ c.m.\ frame. The pion angle in the 
$\gamma N$ c.\ m.\ frame is denoted by $\theta_{\pi}^{\,\star}$. 

For $\pi^-$ production the small reduction at forward angles leads at
350 MeV to an improved and satisfactory description of the data at 
forward angles. At the lower energy of 250 MeV FSI effects are small, 
but one notes in the maximum around 90$^\circ$ an underestimation of the 
data by about 15 percent while at 180$^\circ$ the theory is slightly
higher than  the data. Also at 420 MeV the theory is at forward angles
slightly above the data but below in the backward direction. However,
the overall description is quite satisfactory. The comparison between
theory and experiment is even better for $\pi^0$ production, where the
inclusion of FSI yields an almost perfect description. 

\begin{figure}[htb]
\includegraphics[scale=.7]{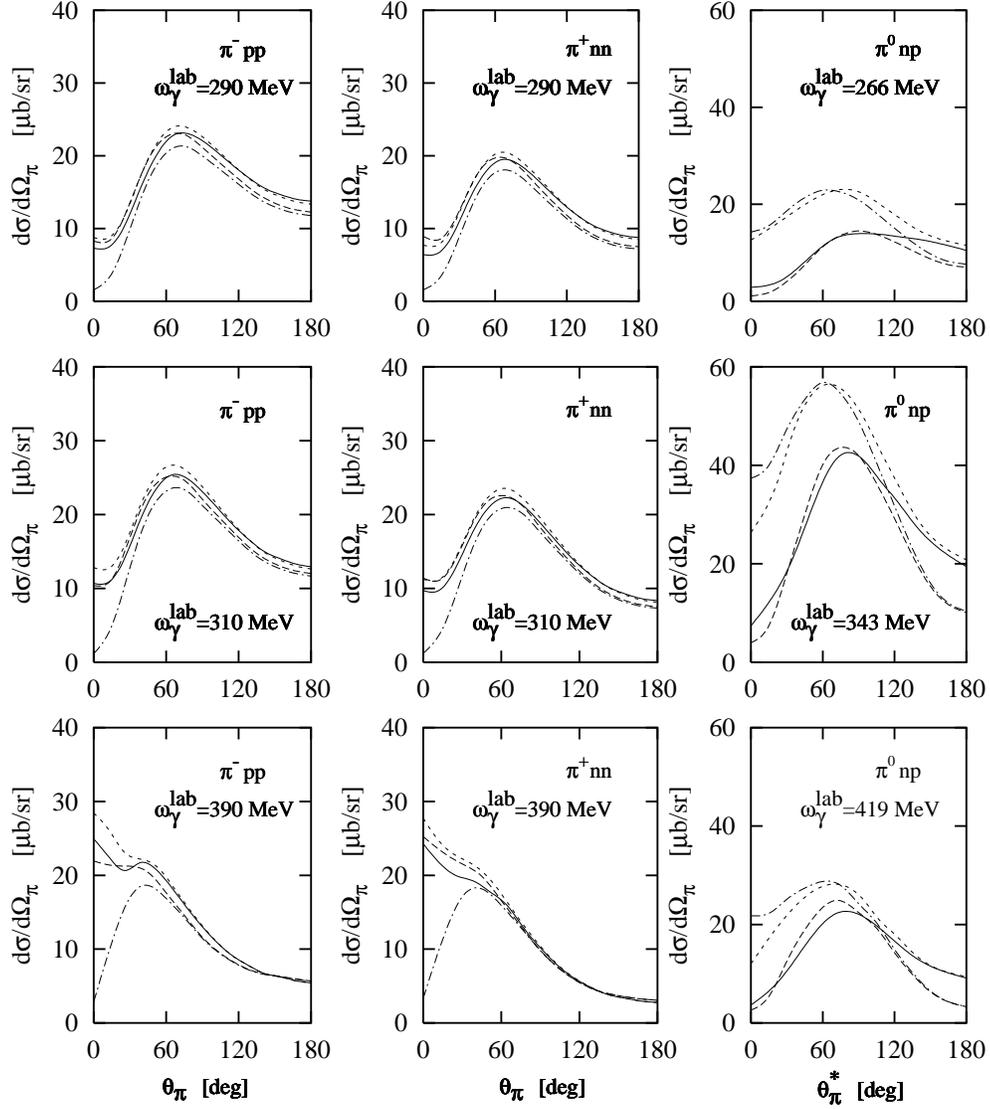}
\caption{Differential cross sections for pion photoproduction on the
  deuteron in comparison with the results from \protect\cite{LeS00} at
  different photon energies. Solid 
  curves: our results for IA plus $NN$ and $\pi N$ rescattering; 
  short-dashed curves: our results in IA; dashed and dash-dotted curves: 
  results from \protect\cite{LeS00} with and without 
  rescattering effects, respectively.}
\label{diffwiththeory}
\end{figure}

\begin{figure}[htb]
\includegraphics[scale=.7]{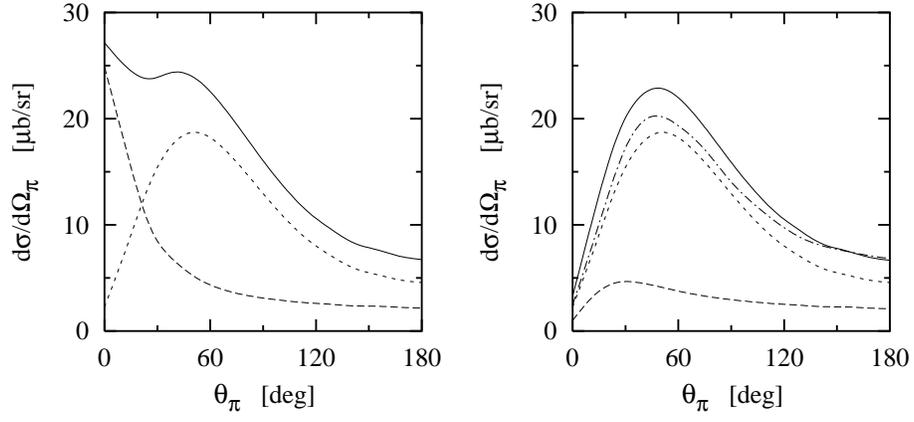}
\caption{Differential cross section for $\pi^-$
photoproduction on the deuteron in the impulse approximation at 
a photon energy of 370 MeV. Left panel: our result with separate 
$s=0$ (dashed curve) and $s=1$ (short-dashed curve) contributions; solid 
curve: total result; Right panel: our result assuming a wrong 
$NN$ antisymmetrization for the $s=0$ channel 
(see (\protect{\ref{nn_antiwrongd}})). Notation as in 
the left panel. In addition the result of \protect\cite{LeS00} 
(dash-dotted curve).} 
\label{wronglev}
\end{figure}

Now we compare our results for the differential cross sections 
with the theoretical predictions of Levchuk {\it et al.}~\cite{LeS00}
in Fig.~\ref{diffwiththeory}. For the $\pi^0$ channel 
we find quite a good agreement in the maximum. At forward angles 
one notes for the pure IA as well as with inclusion of FSI only at the
lowest energy agreement, whereas for the two higher energies larger
differences appear. In the backward direction we find for all three
energies a significantly larger cross section already for the IA while
FSI effects are tiny (see right panels of
Fig.~\ref{diffwiththeory}). Again we suspect  
differences in the elementary production amplitude to be responsible 
for this fact. However, for charged pion production the situation is 
quite different. Major discrepancies are evident in the forward 
direction for the IA. While we find an increased forward peaking of 
the cross section with increasing energy, the cross section remains 
small in~\cite{LeS00} at $\theta_\pi=0$. Analysing in detail the 
contributions from the $s=0$ and $s=1$ parts of the $NN$ final state 
plane wave (see left panel of Fig.~\ref{wronglev}), 
we discovered that we could reproduce the results 
of~\cite{LeS00} if we assume a wrong antisymmetrization for the $s=0$ 
channel, i.e., using instead of (\ref{plane_wave}) for $s=0$
\begin{equation}
|\vec{p}_{1}\vec{p}_{2},0 0,t -\mu \rangle =
  \frac{1}{\sqrt{2}}\left(
|\vec{p}_{1}\rangle^{(1)}|\vec{p}_{2}\rangle^{(2)} + (-)^{t}
|\vec{p}_{2}\rangle^{(1)}|\vec{p}_{1}\rangle^{(2)}\right)|00,t -\mu\rangle\,,
\label{nn_antiwrongd}
\end{equation}
and keeping the form of (\ref{plane_wave}) for the $s=1$ channel. 
This is demonstrated in the right panel of 
Fig.~\ref{wronglev} where we obtain in this 
case also a decrease of the cross section at 0$^\circ$ very similar 
to~\cite{LeS00}. This wrong 
antisymmetrization for the $s=0$ channel corresponds in the uncoupled 
representation, as used in~\cite{LeS00}, to an interchange of the 
momenta of the two nucleons alone without interchanging the spin 
quantum numbers. This we have checked by using also an uncoupled, i.e.,
helicity basis leading to the same result. However, we can only suspect 
that the difference to the results of~\cite{LeS00} may originate from 
such an error in the antisymmetrization.

\section{Summary and conclusions}
\label{sec6}
In this work we have investigated the influence of
final state interaction effects on incoherent single pion 
photoproduction on the deuteron in the $\Delta (1232)$ resonance region. 
The elementary production operator on the nucleon 
is taken in an effective Lagarangian 
model used earlier in a study of the same process in the impulse 
approximation, where all kind of final
state interactions and other two-body operators were neglected. 
As presumably dominant final state interaction effects we have
included the complete rescattering contributions in the two-body 
$NN$- and $\pi N$-subsystems. As models
for the interaction of the $NN$- and $\pi N$-subsystems we used separable
representations of realistic $NN$ and $\pi N$ interactions which 
give a good description of the corresponding phase shifts. 
For $NN$ rescattering, 
we have included all partial waves with total angular momentum $J\le 3$ 
and for $\pi N$ rescattering $S$- through $D$-waves. 

We found that the influence of $NN$ and $\pi N$
rescattering reduces the total cross sections in the $\Delta$(1232) 
resonance region for charged pion photoproduction by about
5 percent and for $\pi^0$ photoproduction reaction, where rescattering is much
more important, by about 35 percent in the maximum. Furthermore, 
$\pi N$ rescattering appears to be much less important compared to $NN$
rescattering. In comparison with experimental data, the inclusion of 
rescattering effects leads to an improved agreement with experimental data. 
Only in the maximum of $\pi^0$ production our model overestimates 
slightly the measured total cross section by a few percent. With respect to the
theoretical predictions of~\cite{LeS00}, we obtained very similar 
results when FSI is included.

The study of the differential cross section revealed that the reduction by
inclusion of FSI appears predominantly at pion
forward angles by about 15 percent for charged pion production and for the  
$\pi^0$ channel by about 40 percent. For pions emitted in the backward
direction the influence of rescattering is much less important. 
As already noted for the total cross section, $\pi N$ rescattering 
has a very small effect on the final results. In comparison with experiment, 
the inclusion of FSI yields a very satisfactory agreement with data. 
Small discrepancies remain at backward pion angles. 
In comparison with the results of \cite{LeS00} in the IA, we found
for charged pion channels at forward angles a large difference between 
both calculations. A detailed analysis gave as possible explanation 
a wrong antisymmetrization for the final two-proton or two-neutron state. 
After inclusion of rescattering effects we obtained a
satisfactory agreement with~\cite{LeS00}.

The present study will serve as a basis for 
further investigations including a three-body treatment the final 
$\pi NN$ system for the lowest and most important partial waves. This
will insure the important unitarity condition and may result in an
even better agreement between  
experimental data and theoretical predictions for $\pi^0$ photoproduction. 
A further interesting topic concerns the study of polarization observables 
giving more detailed information on the dynamics and thus providing more
stringent tests for theoretical models. 
As future refinements we consider also the use of a more 
sophisticated elementary production operator, which will allow one to
extend the present approach to higher energies, and the role of
irreducible two-body contributions to the e.m.\ pion production
operator, e.g.\ interaction of the intermediate particle of the
nucleon and $\Delta$ pole diagrams with the spectator nucleon. In the
long run, one would also need to extend the formalism 
to the threshold region for which the elementary production operator
has to be improved. This process is of great interest since
experimental data for the reaction $d(\gamma,\pi^0n)p$ have been
measured recently in Mainz (MAMI/TAPS) and Saskatoon (SAL)
\cite{Hor01}. Moreover, the formalism should be extended to
investigate coherent and incoherent electroproduction of pions on the
deuteron including final state interaction effects in both the
threshold and the $\Delta$(1232) resonance regions in order to analyze
recent results from MAMI \cite{Mer00}.

\section*{Acknowledgements}
We would like to thank Alexander Fix for useful discussions and 
a critical reading of the manuscript. 
E.M.\ Darwish acknowledges a fellowship from Deutscher 
Akademischer Austauschdienst (DAAD) and would like to thank the
Institut f\"ur Kernphysik of the Johannes Gutenberg-Universit\"at, 
Mainz for the very kind hospitality.
This work was supported by the Deutsche Forschungsgemeinschaft
(SFB 443).

\end{document}